\documentclass[dvips,preprint]{imsart}

\RequirePackage[utf8]{inputenc}
\RequirePackage[OT1]{fontenc}
\RequirePackage[numbers]{natbib}
\RequirePackage{hypernat}
\RequirePackage{bibentry}
\RequirePackage[colorlinks,citecolor=blue,urlcolor=blue]{hyperref}
\RequirePackage{graphics}
\RequirePackage{amsthm,amsmath}
\RequirePackage{booktabs}
\RequirePackage{array}
\RequirePackage{multirow}

\startlocaldefs
\numberwithin{equation}{section}
\theoremstyle{plain}
\newtheorem{thm}{Theorem}[section]
\newtheorem{cor}{Corallary}[section]
\newtheorem{cond}{Condition}[section]
\def\P{{{\rm I}\kern-.18em{\rm P}}}
\def\E{{{\rm I}\kern-.18em{\rm E}}}
\def\R{{{\rm I}\kern-.18em{\rm R}}}
\def\F{{I}\kern-.3em{F}}

\def\J#1#2#3{\mathrm{J}\kern-.18em{\mathrm{I}}_{#1}\left[#2\right]\kern-.1em(#3)}
\def\Jp#1#2#3{\mathrm{J}\kern-.18em{\mathrm{I}}_{#1}\left[#2\right]'\kern-.1em(#3)}
\def\1{{{\mathrm{1}}\kern-.18em{\mathrm{I}}}}
\def\var#1{{\mathrm{var\left[#1\right]}}}
\def\mse#1{{\mathrm{mse\left[#1\right]}}}

\def\aslim{\mathop{{\mathrm{lim}}_{a.s.}}}
\def\e{{\mathrm{e}}}
\newcommand{\convind}{\buildrel{\cal D}\over\longrightarrow}
\endlocaldefs

\begin{document}

\begin{frontmatter}
\title{Estimation of the relative risk following group sequential procedure based upon the weighted log-rank statistic\thanksref{t1}}
\runtitle{Estimation following group sequential trial}
\thankstext{t1}{This article is a U.S. Government work and is in the public domain in the U.S.A.}


\begin{aug}
\author{\fnms{Grant} \snm{Izmirlian}\corref{}\ead[label=e1]{izmirlig@mail.nih.gov}}
\address{National Cancer Institute; Executive Plaza North, Suite 3131\\
6130 Executive Blvd, MSC 7354; Bethesda, MD 20892-7354\\ \printead{e1}}

\runauthor{G. Izmirlian}
\end{aug}

\begin{abstract}
In this paper we consider a group sequentially monitored trial on a survival endpoint, monitored using a weighted
log-rank (WLR) statistic with deterministic weight function. We introduce a summary statistic in the form of a weighted
average logged relative risk and show that if there is no sign change in the instantaneous logged relative risk, there
always exists a bijection between the WLR statistic and the weighted average logged relative risk. We show that this
bijection can be consistently estimated at each analysis under a suitable shape assumption, for which we have listed two
possibilities. We indicate how to derive a design-adjusted p-value and confidence interval and suggest how to apply the
bias-correction method. Finally, we document several decisions made in the design of the NLST interim analysis plan and
in reporting its results on the primary endpoint.   

\end{abstract}





\end{frontmatter}

\section{Introduction}
Time to event, e.g. disease specific mortality, is the primary endpoint in many clinical trials. The use of group
sequential boundaries in monitoring the trial is not only commonplace, but ethically mandated in all trials of human
subjects. The logrank statistic is often the monitoring statistic of choice due to its natural connection with the
relative risk, which is often the parameter of inference.  This natural connection, which is based upon the assumption
of proportional hazards, admits a one-to-one correspondence between the inferential procedure based upon the usual
standard normal scale and that based on the scale of the natural parameter. However, the assumption of proportional
hazards is not always a reasonable assumption. In many subject areas, e.g. in disease-prevention trials, one expects
that the hazard ratio will not be constant.  Much of the prior work on the use of the weighted logrank statistic in a
sequential design is confined to the use a weighting function from the $G^{\rho, \gamma}(t) = S^{\rho}(t)
(1-S(t))^{\gamma}$ family, of Fleming and Harrington, \cite{FlemingT:1991}. They suggest two major types of problems
which can arise. First, they argue that use of the weighted logrank statistic does not reproduce the single point
analysis in the way that is desired. Most notably, they argue, there is no clinically meaningful parameter that allows
the values of the monitoring statistic and sequential boundaries to be cast into a clinically meaningful scale. They
believe that this problem is further aggrivated when the range of the weighting function over the duration of the trial
is quite large, such as is the case with the $G^{0, 1}$ weight function (Gillen and Emerson, \cite{Gillen:2005a}) and
suggest a re-weighting scheme whereby the most weight is given to the most recent data collected at each
analysis. Secondly, they argue that if the chosen weighting function is non-deterministic or trial-specific then it is
impossible to compare results from different clinical trials, (Gillen and Emerson, \cite{Gillen:2005b,
  Gillen:2007}). While the bulk of these cautious remarks are useful to know in their own right, several important
points have been omitted from the discussion.  Firstly, as we will show, there is a natural, clinically meaningful
parameter, the weighted average logged relative risk, that is connected bijectively to the weighted logrank statistic
when there is no change in sign in the instantaneous logged relative risk.  Under suitable shape assumptions, the
bijection can be estimated at each analysis. We will show that the asymptotic distribution of the WLR statistic,
suitably normalized is a Brownian motion plus drift under nothing but boundeness conditions. In two corollaries, we
demonstrate how each of two presented shape assumptions translates into a form of the drift function and consequently,
into an estimator of the weighted average logged relative risk. We then demonstrate how the usual results concerning
monitoring and end of trial estimation follow. Finally, we note that this bijection between the weighted logrank
statistic and the weighted average logged relative risk allows the values of the monitoring statistic, efficacy and
futility boundaries, and reported point estimate and confidence interval to be cast into a clinically meaningful scale.

\section{Terminology and framework}
We consider a two armed randomized trial of the effect of an intervention upon a time to event that is run until
time $\tau$. Let $\tilde T_i$ be the possibly unobserved time to event and let $C_i$ a right censoring time. We assume
non-informative censoring for simplicity. Let $T_i = \tilde T_i \wedge C_i$ be the observed time on study and let
$\delta_i = I(\tilde T_i \leq C_i)$ be the event indicator.  Let $X_i$ indicates membership in the intervention arm
($X_i=1$) or control arm ($X_i=0$). We assume, conditional upon $X_i$, that individuals, $i = 1,\ldots,n$ are
distributed independently and identically. Let $dH_0(t)$ and $dH_1(t)$ be the trial arm specific cumulative hazard
increments. We assume throughout that $H_0(t)$ is finite for all $t$ on $[0,\tau]$. For the instantaneous logged hazard
ratio, we write 
\begin{equation}
\beta(t) = \log \left\{ \frac{dH_1(t)}{dH_0(t)}\right\}\,.
\end{equation}
Let $N_i(t) = I( T_i \leq t, \delta_i = 1)$ and $dN_i(t) = N_i(t) - N_i(t-)$ be the subject level counting 
process and its increments, respectively. Let $N_n(t) = \sum_i N_i(t)$ and $dN_n(t) = N_n(t) - N_n(t-)$ be the
aggregated counting process and its increments, respectively. Note that the following difference is a compensated
counting process martingale:  
\begin{equation}
dM_i(t) = dN_i(t) - I(T_i \geq t) \exp(X_i \beta(t)) dH_0(t)
\end{equation}
Let $E_n(t, 0) = \sum_i X_i I(T_i \geq t)/\sum_i I(T_i \geq t)$ denote the proportion of the population at 
risk at time $t$ in the intervention arm, and let $e(t, 0) = \aslim_{n\rightarrow\infty} E_n(t, 0)$ and
let $G(t) = \aslim dN_n(t)/n$. Let $\F_n(t) = \int_0^t E_n(\xi, 0) (1-E_n(\xi, 0))\,dN_n(\xi)/n$ and let 
$\F(t) = \int_0^t e(\xi, 0) (1-e(\xi, 0))\,dG(\xi)$. We introduce the following notation for cross moment integrals
against $d\F$ over $(0,t)$: 
\begin{equation}
\langle \psi_1 | \F | \psi_2\rangle_t = \int_0^t \psi_1(\xi) \,\psi_2(\xi) d\F(\xi)\,.\label{eqn:brkt}
\end{equation}

\noindent For reasons that will become clear below, we consider the target of our investigation to be the following weighted
average logged relative risk: 
\begin{equation}
\beta^{\star} = \frac{\langle Q | \F |\beta\rangle_{\tau}}{\langle Q | \F | 1 \rangle_{\tau}} \,.\label{eqn:betastar}
\end{equation}
Let $q(t) = \beta(t)/\beta^{\star}$. This provides a representaton of the instantaneous logged relative risk function,
$\beta(t) = \beta^{\star}\,q(t)$ as the product of its weighted average value, $\beta^{\star}$ times a shape function,
$q$. Note it follows that the shape function has weighted average value equal to 1:
\begin{equation}
1 = \frac{\langle Q | \F | q \rangle_{\tau}}{\langle Q | \F | 1 \rangle_{\tau}} \,.\label{eqn:intQqeq1}
\end{equation}

\noindent At follow-up time $t$, the $\sqrt{n}$ normalized score statistic with weighting function $Q$ is:
\begin{equation}
U_n(t) = \frac{1}{\sqrt{n}} \sum_{i=1}^n \int_0^{t} Q(\xi) \left\{ X_i - E_n(\xi, 0)\right\} dN_i(\xi)\,.
\end{equation}
Its estimated variance is:
\begin{equation}
V_n(t) = \frac{1}{n} \int_0^{t} Q^2(\xi) E_n(\xi, 0) \left(1-E_n(\xi, 0)\right) dN_n(\xi) = \langle Q | \F_n | Q \rangle_t\,.
\end{equation}
Let $v(t) = \aslim V_n(t)$. Note that $v(t)= \langle Q | \F | Q \rangle_t$. Let $f_n(t;\tau) = V_n(t)/V_n(\tau)$ and
$f(t; \tau) = v(t)/v(\tau)$.  We will on occasion use the shorthand $f_{n,j}$ and $f_j$ for $f_n(t; \tau)$ and
$f(t;\tau)$, respectively. Also, let $m_n(t) = \langle Q | \F_n | Q \rangle_t$ and $m(t) = \langle Q | \F | Q
\rangle_t$. We consider the weighted log-rank (WLR) statistic at time $t$ on several ``scales'' 
\begin{itemize}
\item[(i)]{The standard normal scale: $Z_n(t) = U_n(t)/\sqrt{V_n(t)}$}
\item[(ii)]{The ``Brownian scale'': $X_n(t) = U_n(t)/\sqrt{V_n(\tau)}$}
\end{itemize}

\section{Main Result}
\begin{cond}\label{cond:betabdd}
The instantaneous logged relative risk function, $\beta$, is bounded on $[0,\tau]$.
\end{cond}

\begin{cond}\label{cond:Qbdd}
The chosen weighting function, $Q$, is bounded on $[0,\tau]$ and deterministic.
\end{cond}
Recall that a weighting functions is always non-negative. The stipulated boundedness in conditions \ref{cond:betabdd} and
\ref{cond:Qbdd} above can be relaxed to being of class $L^2$ with respect to the measure $d\F$, as this is all that is 
really required. 

\noindent While the context will involve monitoring the statistic at a sequence of interim analyses, for the time being,
we suppress this aspect and consider instead the following more general and generic result which holds
under the weakest set of assumptions:
\begin{thm}\label{thm:asymp}
Under conditions \ref{cond:betabdd} and \ref{cond:Qbdd}, then under the family of local alternatives,
$\beta_n^{\star} = b^{\star}/\sqrt{n}$, the score statistic, normalized to the ``Brownian scale'' is asymptotically 
a Brownian motion on $[0, 1]$ plus a drift. 
\begin{equation}
X_n(t) \convind W(f(t;\tau)) +  \mu(t)\,
\end{equation}
where the ``time scale'' for the Brownian motion is the variance ratio or information fraction, $f(t;\tau) = v(t)/v(\tau)$,
and the drift, parameterized by $t$ is   
\begin{equation}
\mu(t) = \frac{\langle Q | \F | q\rangle_t}{\sqrt{\langle Q | \F | Q \rangle_{\tau}}} \, b^{\star}\,.\label{eqn:mut}
\end{equation}
\end{thm}
The proof of \ref{thm:asymp} is given in appendix \ref{sec:thm1proof}. Notice, first, that from equations
\ref{eqn:intQqeq1} and \ref{eqn:mut}, it follows that the value of the drift function at the scheduled end of the trial is
\begin{equation}
\mu(\tau) = \frac{\langle Q | \F | 1\rangle_{\tau}}{\sqrt{\langle Q | \F | Q \rangle_{\tau}}} \, b^{\star}\,.\label{eqn:mutau}
\end{equation}
Thus, without any additional assumptions on the shape function, $q$, we have the following corollary:
\begin{cor}\label{cor:bstarTau}
At the planned conclusion of the trial, $\tau$, an estimate of $\beta^{\star}$ is given by the following:
\begin{equation}
{\widehat \beta^{\star}} = X_n(\tau) \frac{\sqrt{\langle Q | \F_n | Q \rangle_{\tau}}}{\sqrt{n} \,\langle Q | \F_n | 1 \rangle_{\tau}}\,.
\end{equation}
\begin{itemize}
\item[(i)]{${\widehat \beta^{\star}}$ is unbiased}
\item[(ii)]{An estimate of its variance is given by 
  \begin{equation}
  \var{\widehat \beta^{\star}} = \frac{\langle Q | \F_n | Q \rangle_{\tau}}{n \,\langle Q | \F_n | 1 \rangle_{\tau}^2}\,.\label{eqn:varhatbetatau}
  \end{equation}
}
\end{itemize}
\end{cor}

\section{Estimates of $\beta^{\star}$ in a Trial Stopped Early}
\noindent Obtaining an estimate of $\beta^{\star}$ at a trial stopped early due to an efficacy boundary crossing will
require more assumptions on the shape function, $q$. At a minimum in order to have a monotone drift function which is
necessary for propper monitoring, we require the following.
\begin{cond}\label{cond:qnonneg}
The shape function, $q$, is non-negative.
\end{cond}
Since the drift's function's dependence on $t$ is through an integral of a non-negative function, we have the following
corollary:
\begin{cor}\label{cor:bstarq}
If conditions \ref{cond:betabdd}, \ref{cond:Qbdd} and \ref{cond:qnonneg} are true then the conclusion of theorem \ref{thm:asymp}
holds and the drift function is monotone increasing or decreasing in $t$, depending upon the sign of $b^{\star}$. 
\end{cor}
Note also that as the inverse of an increasing function is also increasing, the drift function can also be considered a
monotone function of the information fraction. This would, of course, lead to a natural estimate of $\beta^{\star}$ in a
trial stopped early except for the fact that we have no knowledge of $q$.  In order to have a more useful estimator for
$\beta^{\star}$ in trials stopped early, we opt for a semi-parametric model. In the following, we list two
possibilities. The most natural shape condition to impose is true if our choice of weight function was the optimal one
among all possible choices. 
\begin{cond}\label{cond:qeqKQ}
The shape function, $q$, is proportional to our chosen weighting function, $q(t) = K\,Q(t)$.
\end{cond}
Note that as the weighted average of the shape function must equal 1 as in equation \ref{eqn:intQqeq1} it follows that
the constant of proportionality, $K$, must be 
\begin{equation}
K = \frac{\langle Q | \F | 1 \rangle_{\tau}}{\langle Q | \F | Q \rangle_{\tau}} \,.\label{eqn:Kdef}
\end{equation}

\begin{cor}\label{cor:bstarqeqKQ}
If conditions \ref{cond:betabdd}, \ref{cond:Qbdd} and \ref{cond:qeqKQ} are true then  
\begin{itemize}
\item[(i)]{$X_n$ is asymptotically a Brownian motion with a drift that is linear in the information fraction:
  \begin{equation}
  \mu(t) = \frac{\langle Q | \F | 1 \rangle_{\tau}}{\sqrt{\langle Q | \F | Q \rangle_{\tau}}} f(t; \tau) \, b^{\star}\,.\label{eqn:mutKQ}
  \end{equation}
}
\item[(ii)]{If the trial is stopped at an analysis number $J$ at calender time $t_J$ due to an effacacy boundary
  crossing, then we have the following estimate of $\beta^{\star}$
  \begin{equation}
  {\widehat \beta^{\star}} = \frac{X_n(t_J)}{f_n(t_J; \tau)} \,\frac{\sqrt{\langle Q | \F_n | Q
      \rangle_{\tau}}}{\sqrt{n} \,\langle Q | \F_n | 1 \rangle_{\tau}}  
  \end{equation}
}
\item[(iii)]{An estimate of the mean-squared error is given by:
  \begin{equation}
  \mse{\widehat \beta^{\star}} = \frac{\langle Q | \F_n | Q \rangle_{\tau}}{n \,f_n(t_J; \tau)\,\langle Q | \F_n | 1 \rangle_{\tau}^2} 
  \end{equation}
}
\end{itemize}
\end{cor}
Another natural shape condition is true when we have opted for a weighted statistic but the true shape is constant.
\begin{cond}\label{cond:qeq1}
The shape function, $q$, is identically 1.
\end{cond}

\begin{cor}\label{cor:bstarqeq1}
If conditions \ref{cond:betabdd}, \ref{cond:Qbdd} and \ref{cond:qeq1} are true then
\begin{itemize}
\item[(i)]{$X_n$ is asymptotically a Brownian motion the following drift:
  \begin{equation}
  \mu(t) = \frac{\langle Q | \F | 1 \rangle_{\tau}}{\sqrt{\langle Q | \F | Q \rangle_{\tau}}} r(t, \tau) \, b^{\star}\,,\label{eqn:mut1}
  \end{equation}
  where $r(t; \tau) = \langle Q | \F | 1 \rangle_t/\langle Q | \F | 1 \rangle_{\tau}$, which is an increasing function of
  $t$ and takes the values $0$ at $t=0$ and $1$ at $t=\tau$.    
}
\item[(ii)]{If the trial is stopped at an analysis number $J$ at calender at time $t_J$ due to an effacacy boundary
  crossing, then we have the following estimate of $\beta^{\star}$
  \begin{equation}
  {\widehat \beta^{\star}} = \frac{X_n(t_J)}{r_n(t_J; \tau)} \,
       \frac{\sqrt{\langle Q | \F_n | Q \rangle_{\tau}}}{\sqrt{n} \,\langle Q | \F_n | 1 \rangle_{\tau}}\,, 
  \end{equation}
  where $r_n(t; \tau) = \langle Q | \F_n | 1 \rangle_t/\langle Q | \F_n | 1 \rangle_{\tau}$
}
\item[(iii)]{An estimate of the mean-squared error is given by:
  \begin{equation}
  \mse{\widehat \beta^{\star}} = \frac{f_n(t_J;\tau) \,\langle Q | \F_n | Q \rangle_{\tau}}{n \,r_n(t_J;\tau)^2\,
    \langle Q | \F_n | 1 \rangle_{\tau}^2}  
  \end{equation} 
}
\end{itemize}
\end{cor}

\section{Application to Monitoring and Final Reporting in a Clinical Trial}
The relationship between the drift of the WLR statistic and the weighted average logged relative risk parameter provided
by theorem \ref{thm:asymp} and its corallaries can be used in the monitoring and final reporting of a clinical
trial. 
\subsection{Futility Boundary}
Our comments regarding monitoring a trial are made within the context of boundaries constructed using the Lan-Demets
procedure, \cite{LanK:1983}.  Construction of the efficacy boundary is done under the null hypothesis that the drift
function is identically zero and can be done without appealing to the results presented here. If a futility boundary is
specified in the design then under either of the shape assumptions, one can apply the corresponding corollary
\ref{cor:bstarqeqKQ} or corollary \ref{cor:bstarqeq1} to calculate the drift function at each interim analysis which is
required to compute the futility boundary under the Lan-Demets approach \cite{LanK:1983}. Note that the shape assumption
being made must be part of the interim analysis plan design. In the following discussion we will assume that the optimal
weighting shape condition \ref{cond:qeqKQ} was specified in the design so that the discussion focuses on the application
of corollary \ref{cor:bstarqeqKQ}. In this case, $\beta^{\star}$ is the weighted average logged relative risk for which
the study is powered to detect and must also be specified in the interim analysis plan design. The values of
$v(\tau)=\langle Q | \F | Q \rangle_{\tau}$ and $m(\tau)=\langle Q | \F | 1 \rangle_{\tau}$ at the planned termination
of the study, $\tau$, must also be specified in the interim analysis plan design. We demonstrate in appendix
\ref{sec:EOSfunctionals} when the only source of censoring is administrative censoring or other cause mortality,
how these functionals can be projected for a specific choice of weighting function, $Q$, based upon projected values of
the cross-arm pooled cumulative hazard function at several landmark times on study. We remark here that following
consensus, we recommend using a non-binding futility boundary which is constructed after construction of an efficacy
boundary which ignores the existence of the futility boundary. This is preferred to the joint construction of efficacy
and futility boundaries as that approach results in a discounted efficacy criterion. 

\subsection{Prediction at End of Trial}
When the trial is stopped at an efficacy or futility boundary crossing, or at the scheduled end of the trial, 
and if the optimal weighting shape assumption \ref{cond:qeqKQ} was specified in the design, then corollary
\ref{cor:bstarqeqKQ} can be used to convert the value of the WLR statistic on the Brownian scale, $X_n(t_j)$, to an
estimate of the weighted average logged relative risk, $\widehat\beta^{\star}$. 
Therefore, our point estimate is
\begin{equation}
  {\widehat \beta^{\star}} = \frac{X_n(t_j)}{f_{n,j}} \,\frac{\sqrt{\langle Q | \F_n | Q
      \rangle_{\tau}}}{\sqrt{n} \,\langle Q | \F_n | 1 \rangle_{\tau}}  
\end{equation}
We use the values of $v(\tau)=\langle Q | \F | Q \rangle_{\tau}$ and $m(\tau)=\langle Q | \F | 1 \rangle_{\tau}$ which
are specified in the interim analysis plan design. As mentioned above, when it is obtained at an efficacy boundary
crossing, these type of estimates are known to be biased away from the null (see e.g. Liu and Hall,
\cite{LiuA:1999}). The construction of a design-adjusted confidence interval and adjustment of this estimate for the
above mentioned bias are standard results, especially under the optimal weighting shape condition \ref{cond:qeqKQ} which
leads, in corollary \ref{cor:bstarqeqKQ}, to a drift that is linear in the information fraction. For sake of
completeness, we outline below how to compute a design adjusted p-value, construct a design-adjusted confidence interval
and how to calculate the bias adjusted estimate of the weighted average logged relative risk.  All three of these tasks
involve the sampling density under the null hypothesis of the sufficient statistic, $(J, X_n(t_J))$, where $J$ and
$X_n(t_J)$ are the analysis number and the value of the weighted logrank statistic at an efficacy crossing. The sampling
density of $(J, X_n(t_J))$ takes the following form. First, for $j=1$, $\pi((1,x)) = \P\{X_n(t_1) = x\}$.  For $j>1$, 
\begin{eqnarray}
\pi((j, x) \kern-0.75em &;&\kern-0.75em \mathbf{b}_{1:(j-1)}, \mathbf{f}_{1:j}) \label{eqn:psi}\\
&=& \frac{d}{dx} \P_{H_0}\{J=j \mathrm{~and~} X_n(t_{\ell}) < \sqrt{f_{\ell}} b_{\ell}\,,\, \ell=1,\ldots,j-1, X_n(t_j) = x\}\nonumber
\end{eqnarray}
Here $\mathbf{b}_{1:(j-1)}$ is the sequence of efficacy boundary points at all prior analyses and $\mathbf{f}_{1:j}$ is
the sequence of information fractions at all analyses prior and current. In the following $\mathbf{b}_{1:{\ell}}$ and
$\mathbf{f}_{1:{\ell}}$ for $\ell < 1$ denote the empty sequence. The construction and form of this density is reviewed
in appendix \ref{sec:density}. Let   
\begin{equation}
\bar{\Pi}((j,x) ; \mathbf{b}_{1:(j-1)}, \mathbf{f}_{1:j}) = \int_x^{\infty} \pi((j,\xi); \mathbf{b}_{1:(j-1)}, \mathbf{f}_{1:j})
  d\xi \label{eqn:Psibar} 
\end{equation} 
be the joint probability under $\pi$ that $J=j$ and $X_n(t_j)$ is in the right tail $(x, \infty)$. In order to
calculate a p-value and construct a confidence interval which account for the sequential design, we must choose an
ordering of the sample space for the statistic $(J, X_n(t_J))$.  Here we prefer to use the following ordering:
$(j, x) > (k, y)$ if and only if ($j=k$ and $x>y$) or $j<k$. This ordering is applicable when the rejection region is
convex, as is the case with Lan-Demets boundaries constructed using a smooth spending function. The discussion of
the p-value and of the confidence interval is in the setting of symmetric 2-sided boundaries and when sign of the
alternative hypothesis is positive as it is a simple matter to apply these results to the case where the sign of the
alternative hypothesise is negative. 
\vskip0.5truein
\noindent{\bf{P-value}}\hfil\break
Under the ordering given above, the region further away from the null than $(J, X_n(t_J))$ is the union of 
all prior rejection regions with the right tail at $X_n(t_J)$. Thus the design-adjusted or sequential p-value
is: 
\begin{equation}
\bar{\Pi}((J,X_n(t_J)) ; \mathbf{b}_{1:(J-1)}, \mathbf{f}_{1:J}) + \sum_{\ell=1}^{J-1}\bar{\Pi}((\ell,b_{\ell}) ; 
     \mathbf{b}_{1:{\ell-1}}, \mathbf{f}_{1:\ell})\,,
\end{equation}

\vskip0.5truein
\noindent{\bf{Confidence Interval}}\hfil\break
\noindent If the probability of type one error that remained prior to analysis $J$ is $\alpha_{tot} - \alpha_{J-1}$ then
a two sided design-adjusted confidence interval for $\widehat\beta^{\star}$ is derived as follows. If we denote by $x_u$
the solution in $x$ of the equation 
\begin{equation}
\alpha_{tot} - \alpha_{J-1} = \bar{\Pi}((J,x) ; \mathbf{b}_{1:(J-1)}, \mathbf{f}_{1:J}) + 
  \sum_{\ell=1}^{J-1}\bar{\Pi}((\ell,b_{\ell}) ; \mathbf{b}_{1:{\ell-1}}, \mathbf{f}_{1:\ell})\,,
\end{equation}
then the design-adjusted confidence interval is 
\begin{equation}
\widehat\beta^{\star} \pm \frac{x_u}{\sqrt{f_{n,J}}}\sqrt{\mse{\widehat \beta^{\star}}}\,,
\end{equation}
where $\mse{\widehat \beta^{\star}}$ is the estimated mean-squared error of $\widehat\beta^{\star}$ as given in part
(iii) of corollary \ref{cor:bstarqeqKQ}. Note that when the efficacy boundary is one-sided one can still construct 
a 2-sided confidence interval by replacing $\alpha_{tot} - \alpha_{J-1}$ above with 1/2 its value.

\vskip0.5truein
\noindent{\bf{Bias Adjustment}}\hfil\break
\noindent 
As in Liu and Hall, \cite{LiuA:1999}, bias adjustment is done recursively as follows. First, 
\begin{equation}
\widetilde \zeta(1, x) = \frac{x}{f_1}
\end{equation}
Continuing, 
\begin{equation}
\widetilde \zeta(j, x) =  \int_{-\infty}^{\sqrt{f_j} b_j} \widetilde \zeta(j-1, \xi)\,\pi((j-1, \xi); \mathbf{b}_{1:(j-1)},
  \mathbf{f}_{1:(j-1)}) \, \phi_{_{\Delta_j}}(x-\xi) \,d\xi 
\end{equation}

\noindent The bias adjusted estimate, $\widetilde\beta^{\star}$, of the weighted average logged relative risk, 
$\beta^{\star}$, is obtained by replacing $X_n(t_J)/f_{n,J}$ in part (ii) of corollary \ref{cor:bstarqeqKQ} with
$\widetilde\zeta(J, X_n(t_J))$ to obtain the following:
\begin{equation}
\widetilde\beta^{\star} =  \widetilde\zeta(J, X_n(t_J))\,\frac{\sqrt{\langle Q | \F_n | Q \rangle_{\tau}}}{\sqrt{n} \,
      \langle Q | \F_n | 1 \rangle_{\tau}}
\end{equation}
The design-adjusted confidence interval is the same as given above, but now centered about $\widetilde\beta^{\star}$
\begin{equation}
\widetilde\beta^{\star} \pm \frac{x_u}{\sqrt{f_{n,J}}}\sqrt{\mse{\widehat \beta^{\star}}}\,,
\end{equation}

\section{The NLST}
The design of the National Lung Screening Trial (NLST) \cite{NLST:2011} interim analysis plan stipulated a one-sided
efficacy boundary constructed using the Lan-Demets procedure with a total probability of type one error set to 0.05.
The trial had 90\% power to detect a relative risk of 0.79 at a sample size of 25,000 per arm, accounting for
contamination and non-compliance that could attenuate this effect to 0.85.  The trial began randomization on August 5th,
2002 and concluded randomization on April 26th, 2004.  A non-binding futility boundary was used. The drift was derived
under the optimal weighting shape assumption, \ref{cond:qeqKQ}, and incorporated the design alternative $\beta^{\star} =
\log(0.85)$.  Initial estimates of $v(\tau)$ and $m(\tau)$ were posed in the design. These were updated by using a least
squares quadratic curve to project required future values of $H$ as data accumulated. During the run of the trial,
projected values of the end of trial functionals $v(\tau)$ and $m(\tau)$ did not vary more than $\pm 5\%$.  Interim
analyses occured starting in Spring of 2006 and continued annually until the 5th analysis. The 6th analysis occured 6
months after the 5th.  Data on the primary endpoint was backdated roughly 18 months to allow more complete ascertainment
by the endpoint verification team. The efficacy boundary was crossed at the sixth interim analysis, using data backdated
to January 15th 2009.  Data on the primary endpoint was collected only for events occurring through December 31, 2009 so
this was used as the scheduled termination date. The raw estimated weighted logged relative risk and its design-adjusted
confidence interval were derived. The bias adjusted weighted logged relative risk was compared to the raw estimate. As
the raw estimate is asymptotically unbiased, and since the crude risk ratio is the most straightforward and tangible
summary of the trial results, the trial leadership decided to report the crude
risk ratio together with the exponentiated raw estimate's design-adjusted confidence interval.

\section{Discussion}
We have shown that there is a natural clinically meaningful parameter, the weighted average logged relative risk, that
is connected the weighted logrank statistic. When $\beta(t)$ does not change sign, the connection is a bijection. We
have shown that under suitable shape assumptions, this bijection can be estimated at each analysis. We have shown how
this bijection between the weighted logrank statistic and the weighted average logged relative risk allows the values of
the monitoring statistic, efficacy and futility boundaries, and reported point estimate and confidence interval to be
cast into a clinically meaningful scale. We have indicated how to derive a design-adjusted p-value and confidence
interval and how bias adjustment of the estimate may be done using known methods. Finally, we have documented several
decisions made in the design of the NLST interim analysis plan and in reporting its results on the primary endpoint.

\bibliographystyle{imsart-number}
\bibliography{beta-star}

\section{Appendices}
\subsection{Proof of Theorem \ref{thm:asymp}}
\label{sec:thm1proof}
\noindent We follow the usual method of adding and subtracting the differential of the compensator, and thereby express
$U_n$ as a sum of a term that is asymptotically mean zero Gaussian process and a drift function which grows as $\sqrt{n}$.
\begin{eqnarray}
U_n(t) &=& \frac{1}{\sqrt{n}} \sum_{i=1}^n \int_0^{t} Q(\xi) \left\{ X_i - E_n(\xi, 0)\right\} dM_i(\xi) \nonumber\\
 &&+\;\;\frac{1}{\sqrt{n}} \sum_{i=1}^n \int_0^{t} Q(\xi) \left\{ X_i - E_n(\xi, 0)\right\} I(T_i\geq \xi) 
     \exp(X_i q(\xi) \beta^{\star}) dH_0(\xi)\nonumber\\
&&\nonumber\\
&=& \frac{1}{\sqrt{n}} \sum_{i=1}^n \int_0^{t} Q(\xi) \left\{ X_i - E_n(\xi, 0)\right\} dM_i(\xi) \nonumber\\
 &&+\;\;\sqrt{n} \int_0^{t} Q(\xi) \left\{ E_n(\xi, \beta^{\star}) - E_n(\xi, 0)\right\} R_n(\xi, \beta^{\star}) dH_0(\xi)\,,
\end{eqnarray}
where in the above, $R_n(\xi, \beta^{\star}) = 1/n \sum_i I(T_i \geq \xi)\exp(X_i q(\xi) \beta^{\star})$, and 
$E_n(\xi, \beta^{\star}) = 1/(n R_n(\xi, \beta^{\star})) \sum_i X_i I(T_i \geq \xi)\exp(X_i q(\xi) \beta^{\star})$.

\noindent By linearizing the difference, $E_n(\xi, \beta^{\star}) - E_n(\xi, 0)$ about $\beta^{\star} = 0$ we obtain
\begin{eqnarray}
U_n(t) &=& \frac{1}{\sqrt{n}} \sum_{i=1}^n \int_0^{t} Q(\xi) \left\{ X_i - E_n(\xi, 0)\right\} dM_i(\xi) \nonumber\\
 &&+\;\;\sqrt{n} \beta^{\star} \int_0^{t} Q(\xi) q(\xi) E_n(\xi, 0)\left\{1 - E_n(\xi, 0)\right\} R_n(\xi, \beta^{\star}) dH_0(\xi)\,.\nonumber\\
 &&\label{eqn:U}
\end{eqnarray}
We normalize by $\sqrt{V_n(\tau)}$ and replace the differential $R_n(\xi, \beta^{\star}) dH_0(\xi)$ with
$dN_n(\xi)/n$. The latter is possible because integrals of bounded functions against the difference of the differentials
are consistent to zero.
\begin{eqnarray} 
X_n(t) &=& \frac{1}{\sqrt{n\,V_n(\tau)}} \sum_{i=1}^n \int_0^{t} Q(\xi) \left\{ X_i - E_n(\xi, 0)\right\} dM_i(\xi) \nonumber\\
 &&+\;\;\sqrt{\frac{n}{V_n(\tau)}} \beta^{\star} \int_0^{t} Q(\xi) q(\xi) E_n(\xi, 0)\left\{1 - E_n(\xi, 0)\right\} \frac{dN_n(\xi)}{n}\nonumber\\
 &&\nonumber\\
&=& W_n(f_n(t;\tau)) \;+\; \frac{\langle Q | \F_n | q\rangle_t}{\sqrt{\langle Q | \F_n | Q \rangle_{\tau}}} \, \sqrt{n} \beta^{\star}\,.
\end{eqnarray}
The first term is easily recognized to be asymptotic in distribution to a standard Brownian motion. The reader can
either directly apply Robolledo's martingale central limit theorem, verifying that in the case that integrands and
intensities are bounded all conditions are satisfied, or apply a more direct result, such as theorem (6.2.1) in
Fleming and Harrington \cite{FlemingT:1991}. Under the family of local alternatives, $\beta^{\star}_n =
b^{\star}/\sqrt{n}$, then by the comments following expression \ref{eqn:U}, the second term is easily seen to be
consistent to the drift function listed in expression \ref{eqn:mut}. Therefore the result follows by Slutzky's theorem. 

\subsection{End of Trial Functionals}
\label{sec:EOSfunctionals}
\noindent In this section we demonstrate how to project values of the variance $v(\tau) = \langle Q | \F | Q
\rangle_{\tau}$, and the ``first moment'' $m(\tau) = \langle Q | \F | 1 \rangle_{\tau}$ at the scheduled end of study,
$\tau$. This is done in the specific case of the ``ramp plateau'' weighting function which was used for interim
monitoring and reporting in the NLST. This is the function which takes the value 0 at $t=0$, has linear increase to the
value 1 at $t=t_c$ and then maintains this constant value forward.
\begin{equation}
Q(t) = \frac{t}{t_c} \wedge 1
\end{equation}
In the NLST, the value of $t_c=4$ years was used. Next, by imposing some mild assumptions we will be able to express all
quantities in the integrands in terms of the  cross-arm pooled cancer mortality cumulative hazard function, $H$ and
thereby solve the integrals via a simple change of variables. The resulting expressions require only values of $H(t)$ at
$t=t_c$, $t=\tau-t_{er}$ and $t=\tau$, where $t_{er}$ is the calender time at which randomization was concluded.  First
we shall list the required assumptions. In the following discussion, $S$, $S_{lr}$ and $S_{oth}$ are survival functions
corresponding to the cross-arm pooled cancer mortality, administrative censoring or ``live removal'' and other cause
mortality. The latter two were the only sources of censoring in the NLST because complete ascertainment with respect to
mortality was possibly through the use of the matching death certificates through the national death index.  

\begin{cond}\label{cond:prop}
Other cause mortality is proportional to cancer mortality, i.e. that $\theta = -dlog(S_{oth})/dH$ is
constant.
\end{cond}

\begin{cond}\label{cond:propall}
Proportional allocation: $e(\xi, 0) \equiv e(0, 0)$.
\end{cond}

\begin{cond}\label{cond:unifaccrH}
Accrual is uniform on the scale of $H$, so that 
\begin{equation}
S_{lr}(\xi) =  \frac{H(\tau) - H(\xi)}{H(\tau) - H(\tau - t_{er})}\wedge 1,
\end{equation}
where $\tau$ is the time at which the required number of events are obtained,
and $t_{er}$ is the time at which randomization is completed. 
\end{cond}

\begin{cond}\label{cond:wtfn}
\begin{equation}
Q(\xi) = \frac{\xi}{t_c} \wedge 1 \equiv  \frac{1-\exp(-H(\xi) \wedge H(t_c) )}{1-\exp(-H(t_c))}.
\end{equation}
\end{cond}

\noindent The other cause versus cancer proportionality assumption is perhaps the most arguable. However, the extent to
which it is violated in practice has little impact upon our results as other cause mortality enters our results only
through its survival function which maintains a value in excess of 0.95 throughout the trial. The proportional
allocation assumption approximates what we see in practice quite closely, especially in the case of a large trial of a
rare event. In the NLST there was 1 to 1 randomization so that $e(0, 0) = 1/2$.  The extent to which the latter two
assumptions \ref{cond:unifaccrH} and \ref{cond:wtfn} hold both depend upon the extent to which pooled cancer specific
mortality grows at a constant rate. In the case of the NLST, the pooled cancer mortality cumulative hazard function
did grow at an approximately linear rate.
\hfil\break 

\noindent{\bf{Variance at Planned Termination}}
\begin{eqnarray}
v(\tau) &=& \langle Q | \F | Q \rangle_{\tau} = \int_0^{\tau} Q^2(\xi) e(\xi, 0) \left(1-e(\xi, 0)\right) dG(\xi) \nonumber\\
&&\nonumber\\
&=& \int_0^{\tau} Q^2(\xi) e(\xi, 0) \left(1-e(\xi, 0)\right) S_{oth}(\xi) S_{lr}(\xi) S(\xi) dH(\xi)\,.\label{eqn:computeV}
\end{eqnarray}
Here, $S$, $S_{lr}$ and $S_{oth}$ are survival functions corresponding to the cross-arm pooled cancer mortality,
administrative censoring or ``live removal'' and other cause mortality. The latter two were the only sources of
censoring in the NLST because complete ascertainment with respect to mortality was possibly through the use of the
matching death certificates through the national death index. Therefore, we can express the differential, $dG$, in this
way. Under assumptions \ref{cond:prop}, \ref{cond:propall}, \ref{cond:unifaccrH}, and \ref{cond:wtfn}, we apply the
change of variables, $\eta = H(\xi)$, to obtain  
\begin{eqnarray*}
v(\tau) &=& \frac{1}{4} \int_0^{H(\tau)} \left(1-\e^{-(\eta\wedge H(t_c))}\right)^2 \,\e^{-\theta \eta}  
                   \left\{\frac{H(\tau) - \eta}{H(\tau) - H(\tau-t_{er})} \wedge 1 \right\} \,\e^{-\eta} d\eta\\
\\
&=& \frac{1}{4} \int_0^{H(t_c)\wedge H(\tau-t_{er})} \left(1 - 2 \e^{-\eta} +
                   \e^{-2\eta}\right) \,\e^{-(\theta + 1)\eta} d\eta\\
\\
&&\;\;+ \frac{I\left(t_c < \tau - t_{er}\right)}{4}  \,\left(1-\e^{-H(t_c)}\right)^2\,
                   \int_{H(t_c)}^{H(\tau - t_{er})} \e^{-(\theta + 1)\eta} d\eta\\
\\
&&\;\;+ \frac{I(\tau-t_{er} < t_c)}{4\left(H(\tau) - H(\tau-t_{er})\right)} \,
                   \int_{H(\tau-t_{er})}^{H(t_c)}\,\left(1 - 2\e^{-\eta} +
                        \e^{-2\eta}\right)\,\e^{-(\theta+1)\eta}\,\left(H(\tau) - \eta\right) d\eta\\
\\
&&\;\;+ \frac{\left(1-\e^{-H(t_c)}\right)^2}{4\left(H(\tau)-H(\tau-t_{er})\right)}\,
                   \int_{H(\tau-t_{er})\vee H(t_c)}^{H(\tau)} \e^{-(\theta+1)\eta} \, 
                     \left(H(\tau)-\eta\right) d\eta\\
\\
&=& I_1 + I_2 + I_3 + I_4\,.
\end{eqnarray*}

\noindent These evaluate to:
\begin{eqnarray*}
I_1 &=& \frac{1}{4}\left\{\frac{1-\e^{-(\theta+1)H_m}}{\theta+1}\;-\;2\,\frac{1-\e^{-(\theta+2)H_m}}{\theta+2}\;+\;
              \frac{1-\e^{-(\theta+3)H_m}}{\theta+3}\right\}\;\;{\rm where~} H_m = H(t_c)\wedge H(\tau-t_{er})\,,\\
\\
I_2 &=& I(t_c<\tau-t_{er})\,\left(1-\e^{-H(t_c)}\right)^2\,
          \frac{\e^{(\theta+1)H(t_c)}-\e^{-(\theta+1)H(\tau-t_{er})}}{4(\theta+1)}\,,\\
\\
I_3 &=& \frac{I(\tau-t_{er} < t_c)}{4(H(\tau)-H(\tau-t_{er}))}\\
\\
&&\;\; \times \;\left\{
\left(\frac{\e^{-(\theta+1)H(\tau-t_{er})}}{\theta+1}- 2\frac{\e^{-(\theta+2)H(\tau-t_{er})}}{\theta+2} +
      \frac{\e^{-(\theta+3)H(\tau-t_{er})}}{\theta+3}\right)\left(H(\tau)-
      H(\tau-t_{er})\right) \right.\\
\\
&&\qquad-\;
\left(\frac{\e^{-(\theta+1) H(t_c)}}{\theta+1} -
     2\frac{\e^{-(\theta+2) H(t_c)}}{\theta+2} +
      \frac{\e^{-(\theta+3) H(t_c)}}{\theta+3}\right)\left(H(\tau) - H(t_c)\right) \\
\\
&&\qquad-\;
\left(\frac{\e^{-(\theta+1)H(\tau-t_{er})}-\e^{-(\theta+1) H(t_c)}}{(\theta+1)^2} 
   - 2\frac{\e^{-(\theta+2)H(\tau-t_{er})}-\e^{-(\theta+2) H(t_c)}}{(\theta+2)^2}\right. \\
\\  
&&\qquad\qquad\qquad\qquad\qquad\qquad\qquad\qquad\qquad
+\;\left.\left.\frac{\e^{-(\theta+3)H(\tau-t_{er})}-\e^{-(\theta+3) H(t_c)}}{(\theta+3)^2}\right)\right\}\,,\\ 
\\
I_4 &=& \frac{\left(1-\e^{-H(t_c)}\right)^2}{4 (\theta+1)}\\
\\
&&\kern-1em\times \;\left\{
\frac{H(\tau) -(H(\tau-t_{er})\vee H(t_c))}{H(\tau)-H(\tau-t_{er})}\,\e^{-(\theta+1)\left(H(\tau-t_{er}) \vee H(t_c) \right)}\;
-\;\frac{\e^{-(\theta+1)(H(\tau-t_{er}) \vee H(t_c))}-\e^{-(\theta+1)H(\tau)}}{(\theta+1)(H(\tau)-H(\tau-t_{er})}\right\}
\end{eqnarray*}
respectively.
\vskip0.5truein
\noindent{\bf{First Moment at Planned Termination}}
\begin{eqnarray}
m(\tau)  &=& \int_0^{\tau} Q(\xi) e(\xi, 0) \left(1-e(\xi, 0)\right) dG(\xi) \nonumber\\
&&\nonumber\\
&=& \int_0^{\tau} Q(\xi) e(\xi, 0) \left(1-e(\xi, 0)\right) S_{oth}(\xi) S_{lr}(\xi) S(\xi) dH(\xi)\,.\label{eqn:computeM}
\end{eqnarray}

\noindent Under assumptions \ref{cond:prop}, \ref{cond:propall}, \ref{cond:unifaccrH}, and \ref{cond:wtfn}, we again
apply the change of variables, $\eta = H(\xi)$, to obtain  
\begin{eqnarray*}
m(\tau) &=& \frac{1}{4} \int_0^{H(\tau)} \left(1 - \e^{-\eta\wedge H(t_c)}\right)\,\e^{-\theta
  \eta}\,\left\{\frac{H(\tau) - \eta}{H(\tau) - H(\tau-t_{er})} \wedge 1 \right\} \,\e^{-\eta} d\eta\\
\\
&=& \frac{1}{4} \int_0^{H(t_c)\wedge H(\tau-t_{er})} \left(1-\e^{-\eta}\right)\,\e^{-\theta \eta}\,\e^{-\eta} d\eta\\
\\
&&\;+\;\frac{1}{4}\,I(t_c < \tau-t_{er})\,\left(1-\e^{-H(t_c)}\right)\,\int_{H(t_c)}^{H(\tau-t_{er})}\,\e^{-\theta\eta}\,
  \e^{-\eta} d\eta\\
\\
&&\;+\; \frac{1}{4}\,I(t_c > \tau-t_{er})\,\int_{H(\tau-t_{er})}^{H(t_c)} \left(1-\e^{-\eta}\right)\,\e^{-\theta \eta}\,
    \frac{H(\tau) - \eta}{H(\tau) - H(\tau-t_{er})}\,\e^{-\eta} d\eta\\
\\
&&\;+\; \frac{1}{4}\,I(t_c < \tau) \,\left(1-\e^{-H(t_c)}\right)\,\int_{H(t_c)\vee H(\tau-t_{er})}^{H(\tau)}\,
    \e^{-\theta \eta}\,\frac{H(\tau) - \eta}{H(\tau) - H(\tau-t_{er})} \,\e^{-\eta} d\eta\\
\\
&=& J_1 + J_2 + J_3 + J_4
\end{eqnarray*}

\noindent These evaluate to
\begin{eqnarray*}
J_1 &=& \frac{1}{4} \left\{\frac{1-\e^{-(\theta+1)(H(t_c)\wedge H(\tau-t_{er}))}}{\theta+1} -
    \frac{1-\e^{-(\theta+2)(H(t_c)\wedge H(\tau-t_{er}))}}{\theta+2} \right\}\,,\\
\\
J_2 &=& \frac{1}{4}\,I(t_c < \tau-t_{er})\,\left(1-\e^{-H(t_c)}\right)\,
     \frac{\e^{-(\theta+1)H(t_c)} - \e^{-(\theta+1)H(\tau-t_{er})}}{\theta+1}\,,\\
\\
J_3 &=&\frac{I(t_c > \tau-t_{er})}{4\left(H(\tau) - H(\tau-t_{er})\right)}\\
\\
&&\qquad\qquad\times\;\left\{\left(\frac{\left(H(\tau)-H(\tau-t_{er})\right)\,\e^{-(\theta+1)H(\tau-t_{er})} - 
                       \left(H(\tau)-H(t_c)\right)\,\e^{-(\theta+1)H(t_c)}}{\theta+1}\right.\right.\\
\\
&&\qquad\qquad -\left.\frac{\left(H(\tau)-H(\tau-t_{er})\right)\,\e^{-(\theta+2)H(\tau-t_{er})} - 
                       \left(H(\tau)-H(t_c)\right)\,\e^{-(\theta+2)H(t_c)}}{\theta+2}\right)\\
\\
&&\qquad\qquad -\left.\left(\frac{ \e^{-(\theta+1)H(\tau-t_{er})} - \e^{-(\theta+1)H(t_c)}}{(\theta+1)^2}
        \;-\;\frac{\e^{-(\theta+2)H(\tau-t_{er})} - \e^{-(\theta+2)H(t_c)}}{(\theta+2)^2}\right)\right\}\\
\\
J_4 &=&\frac{I(t_c < \tau) \,\left(1-\e^{-H(t_c)}\right)}{4\left(H(\tau) - H(\tau-t_{er})\right)}\\
\\
&&\quad\times\;\left\{\frac{\left(H(\tau)-H(t_c\vee(\tau-t_{er}))\right)\,
                          \e^{-(\theta+1)H(t_c\vee(\tau-t_{er}))}}{\theta+1}
\; - \; \frac{\e^{-(\theta+1)H(t_c\vee(\tau-t_{er}))} - \e^{-(\theta+1)H(\tau)}}{(\theta+1)^2}\right\}
\end{eqnarray*}
respectively.
\vskip0.5truein
\noindent{\bf{Duration of Trial}}\hfil\break
\noindent The duration the NLST was part of the design. In other situations in which the design stipulates that the trial should
run until required number of events is attained, the above change of variables technique can be used to 
find a closed form expression for 
\begin{equation}
G(\tau) = \int_0^{\tau} S_{oth}(\xi) S_{lr}(\xi) S(\xi) dH(\xi)\,.\label{eqn:computeG}
\end{equation}
in terms of the projected values of $H$ at $t=\tau$ and $t=\tau-t_{er}$.  Then using the plug-in estimate $\E
N_n(\tau)/n$ for $G(\tau)$ this expression can be inverted to solve for $\tau$, the duration of the trial.

\subsection{Sampling density of $(J, X_n(t_J))$}
\label{sec:density}
As in Armitage, McPherson and Rowe, \cite{ArmitageP:1969}, the sampling density of $(J, X_n(t_J))$ can be derived
recursively as follows. Let $\Delta_j = f_{n,j} - f_{n,j-1}$ and let $\phi_v(x) = \phi(x/\sqrt{v})/\sqrt{v}$ where $\phi$ is
the density of the standard normal. First,  
\begin{eqnarray}
\pi((1,x)) = \phi_{_{f_1}}(x)\,.
\end{eqnarray}
Next, for all $j> 1$, 
\begin{eqnarray}
\pi((j,x) \kern-1em &;&\kern-1em \mathbf{b}_{1:(j-1)}, \mathbf{f}_{1:j}) \nonumber\\
&& \nonumber\\
\kern-1em&=&\kern-1em \int_{-\infty}^{\sqrt{f_{j-1}} b_{j-1}} \pi((j-1, \xi);\mathbf{b}_{1:(j-2)},\mathbf{f}_{1:(j-1)})\,
  \phi_{_{\Delta_j}}(x-\xi) \,d\xi \nonumber\\
 \label{eqn:pij}
\end{eqnarray}

\end{document}